\documentclass{JHEP}
\usepackage{epsfig}
\usepackage{graphicx}%
\def\be{\begin{equation}}
\def\ee{\end{equation}}
\def\bea{\begin{eqnarray}}
\def\eea{\end{eqnarray}}
 
\def\lesssim{\mathrel{\hbox{\rlap{\hbox{\lower4pt\hbox{$\sim$}}}\hbox{$<$}}}}
\def\gtrsim{\mathrel{\hbox{\rlap{\hbox{\lower4pt\hbox{$\sim$}}}\hbox{$>$}}}}

\title{Problems  with Tachyon  Inflation}

\author{Lev Kofman\\
    CITA, University of
Toronto, 60 St. George Street, Toronto, ON M5S 3H8, Canada\\ E-mail: 
\email{kofman@cita.utoronto.ca}}
\author{Andrei Linde\\
    Department of Physics, Stanford University, Stanford, CA 94305,
USA\\
    E-mail: \email{alinde@stanford.edu}, ~~ 
http://physics.stanford.edu/linde}
\received{\today}       
 \preprint{CITA-2002-14\\ SU-ITP-02/19\\ \hepth{0205121}\\ May 13, 2002}

\abstract{We consider  cosmological consequences of string theory tachyon condensation. We show that it is very difficult to obtain inflation in the simplest versions of this theory. Typically, inflation in these theories could occur only at super-Planckian densities, where the  effective 4D field theory is inapplicable.  Reheating and creation of matter in  models where
the minimum of the 
 tachyon potential $V(T)$ is  at $T\to \infty$ is problematic because the tachyon field in such theories  does not oscillate. If the universe after inflation is dominated by the energy density of the tachyon condensate, it will
 always remain dominated by the tachyons.   It might happen that string condensation is responsible for a short stage of inflation at a nearly Planckian density, but one would need to have a second stage of inflation after that. This would imply that the tachyon played no role in the post-inflationary universe until the very late stages of its evolution. These problems do not appear in the recently proposed models of hybrid inflation  where the complex tachyon field has a minimum at $T \ll M_p$.
}


\begin{document}

\section{Introduction}\label{introduction}

The string  theory  tachyon \cite{sennonbps},
 in one form or another, appears in various models 
 of brane inflation.
There are two classes of such models. In the first class we will deal with non BPS branes,
which ultimately annihilate or decay. In the second class we will deal with
unstable  configurations of   near-BPS branes, which will be 
dynamically reconfigured into stable states. 
 In the first models of  brane inflation
the  inflaton field was associated with the inter-brane separation \cite{DvaliTye}.
Early  models of this type \cite{stringyhybrid} considered a simple
 system of colliding  and annihilating $D$ and $\bar D$ branes
  giving rise to  hybrid inflation \cite{Hybrid}. At the end of hybrid inflation there is a stage of tachyonic instability \cite{tach}. In  \cite{stringyhybrid} it was speculated that this stage might be associated
 with string tachyon condensation. 
The models of annihilating branes  belong to the first class of
 brane inflation models. 

However, dynamical  theory of brane-antibrane annihilation
and time evolution of tachyon condensate  is rather complicated, and the shape of the tachyon potential
 was based on various conjectures.
Significant progress in  was achieved in   \cite{renata,renata2}
for the second class of brane inflation models,
where instead of  brane-antibrane annihilation the authors considered
BPS  branes either at small angles \cite{renata} or in the presence of small background fields \cite{renata2}. 
This made it possible to calculate the  full effective potential describing
 the system both at the stage of inflation and at the subsequent
stage of tachyon instability. 
  The effective potential of both of these models, including
 radiative corrections,   coincides  with the potential of the  $N=2$ 
supersymmetric model of $P$-term inflation \cite{Kallosh:2001tm}. 
Other interesting models of a similar type  have
 been proposed in \cite{shafi} -\cite{Sarangi:2002yt}.

One may also study cosmological consequences of
 tachyon condensation in the context of the theory of a single unstable D-brane \cite{sennonbps}.  This is another version of the first class of 
brane inflation.
The form of the tachyon potential $V(T)$ may depend on the underlying
 theory (bosonic or supersymmetric).
In particular, for bosonic and supersymmetric string field theory
 studied in \cite{sennonbps,shatash,Kutasov1,Kutasov2,Choudhury:2002xu} the
unstable tachyon mode has a ground state at   $T\to \infty$ \cite{Sen1}. 
 In this respect such models considerably differ  from the models
of the second class 
 studied in \cite{renata} - \cite{Sarangi:2002yt}.
 Recently Sen proposed  considering  the tachyon rolling towards its minimum at infinity
as a dark matter candidate \cite{Sen1}.
 Several authors have investigated 
 the process of rolling of the tachyon in the cosmological 
background 
\cite{Gibb,Frolov:2002rr,Shiu:2002qe,Padmanabhan}. 

However, in Section 3 of this paper we will point out that if the field $T$ rolls down from the maximum of its potential and the universe does not inflate, then the quantum fluctuations produced during spontaneous 
symmetry breaking and generation of tachyon condensate divide  the 
universe into many small into domains of 
different types. The resulting structure is incompatible with 
cosmological observations. 

The main goal of our paper is to study the possibility of inflation during  tachyon condensation. This would resolve the problem mentioned above.
String theory motivated 
tachyon inflation was discussed 
 in \cite{Fairbairn:2002yp,Choudhury:2002xu} \footnote{In papers 
\cite{Feinstein:2002aj,Padmanabhan:2002cp} inflation was considered for
phenomenological  potentials $V(T)$ hat have not been derived in the string tachyon theory.  Such models are related to
phenomenological ``k-inflation'' \cite{k}}.
However, these  investigations were rather incomplete.
 In Section 4  we will re-analyse the possibility of 
inflation during  tachyon condensation.
 We will show, contrary to what was claimed in 
\cite{Fairbairn:2002yp,Choudhury:2002xu}, that it is very difficult to find realistic inflationary models in this theory:
 Inflation near the maximum of $V(T)$ requires a strong coupling regime $g_s \gg 1$, which, in its turn,  leads to unacceptably large perturbations of the metric.

In addition to these problems, there is another one. A tachyon rolling to its ground state at $T\to \infty$  
almost immediately acquires a matter dominated  equation of state (no pressure).
Its energy density  decreases  as 
$\varepsilon \sim a^{-3}$, whereas the density of radiation decreases as $a^{-4}$. 
  Therefore if the tachyon originally dominated the energy density of the universe (and it is the case if its energy density was responsible for inflation), then it will {\it always} dominate the energy density of the universe  \cite{Shiu:2002qe}.  
In Section 5 we will argue that this fact   rules out not only inflationary models of the type of
 \cite{Fairbairn:2002yp,Choudhury:2002xu}, but also those versions of hybrid inflation where the tachyon potential has
a  minimum at $T \to \infty$.
 
  This does not mean that the rolling tachyon cannot play the role of dark matter. This means, however, that in those models where the tachyon may serve as a dark matter candidate at the present time, its
energy density after inflation should be  fine tuned to be 
subdominant until the very late stages of the evolution of the universe. 

Thus, we will argue that  tachyon condensation of the type considered in \cite{sennonbps} does not lead to inflation and is not expected to play an important role in  post-inflationary early universe cosmology.

\section{Effective Field Theory of Tachyon Matter }\label{theory}

In this Section we review the field theory of tachyon matter. 
Consider  the 4D effective field theory
 of the  
tachyon field $T$  coupled to Einstein gravity
\begin{equation}\label{action}
S = \frac{M_p^2}{2} \int d^4x\, \sqrt{g}R+S_{B,S}  \ .
\end{equation}
The 4D effective field theory  action  of 
 the  tachyon field $T$  
on a D$3$-brane,
 computed in bosonic theory around the top of the potential,
 up to higher derivative terms, is given by
 \cite{shatash,Kutasov1}
\begin{equation}\label{action1}
S_B =  \tau_3\int d^4x\, \sqrt{g}\left(\alpha'  e^{-T} \partial_\mu{T}\partial^\mu{T}
 + (1+T)  e^{-T}\right) \ .
\end{equation}
Similarly, a
non-BPS D$3$-brane action  in 
supersymmetric (B-SFT) theory \cite{Kutasov2}
\begin{equation}\label{action5}
S_S =  \tau_3\int d^4x\,\left(\alpha' \ln 2\,e^{-{ T}^2/4}\,
\partial_\mu{T}\partial^\mu{T} + e^{-{T}^2/4}\right) \ .
\end{equation}
In this theory the tension of the non-BPS brane is greater than that of $BPS$ branes by a factor of $\sqrt{2}$  \cite{Kutasov2}:
\begin{equation}\label{tension}
\tau_3 = {M_s^4 \sqrt{2}\over {(2\pi)^3 g_s}} \ ,
\end{equation}
where $g_s$ is the string coupling, and $M_s =  l_s^{-1} = 1/\sqrt{\alpha'}$  are the 
fundamental string  mass and length scales. 
The Planck mass in  4D  is obtained by
dimensional reduction. 
According to Ref. \cite{Jones:2002cv},
\begin{equation}\label{planck}
M_p^2=  {M_s^2 v\over g_s^2 } \ ,
\end{equation}
where $v =  (M_s r)^d/\pi$,  $r$ is a radius of the compactification, and $d$ is 
the number of  compactified dimensions. Note that the volume of the
 compactified space is greater than  $(M_s r)^d$ by a factor of $(2\pi)^d$.
 Usually one assumes that $r \gg l_s$, i.e $v \gg 1$, in order to be able to
 use the effective 4D field theory  \cite{Jones:2002cv}.\footnote{In Ref.  \cite{Fairbairn:2002yp} one power of $g_s$ in the expression for $M_p^2$ was missing, which affected their estimates of the magnitude of perturbations of metric.}

It could be useful to obtain closed form expressions for  the action (\ref{action1}), (\ref{action5}) incorporating
all higher powers of $\partial_{\mu}T$.
There are suggestions to extend actions (\ref{action1}), (\ref{action5})
to include all powers of the derivatives $\partial_\mu{T}$.
Effective tachyon field theory in the Born-Infeld-type form  
\cite{Sen1,Gibb} is described by
\begin{equation}\label{action4}
S_{B,S} =  \int d^4x\, \sqrt{g} V(T) \,\sqrt{1+
\alpha' \,\partial_\mu{T}\partial^\mu{T}} \ .
\end{equation}
In this form the effective tachyon field theory appears in
recent papers on tachyon cosmology.

To match (\ref{action4}) with the truncated actions  (\ref{action1}),
(\ref{action5}), one can redefine the tachyon  field $T$.
For bosonic theory we can make a redefinition 
$T \to \sqrt{8(1+T)}$
and   the potential around the maximum becomes
\begin{equation}\label{pot1}
V(T) \simeq \frac{e}{8}\, \tau_3 \,  T^2 \, e^{-\frac{T^2}{8}} \ .
\end{equation}
This potential has a maximum at $ T=\sqrt{8}$ and is defined for $ T \geq 1$.
For supersymmetric theory 
  $ T \to \sqrt{2\ln 2}\,T$ 
and  the potential around the maximum is 
\begin{equation}\label{pot2}
V(T) \simeq \tau_3 \,  e^{-\frac{ T^2}{8\ln 2}} \ .
\end{equation}
This potential has a maximum at $ T=0$ and is symmetric,
$V(-T)=V(+ T)$.

However, at large $T$,
according to Sen \cite{Sen1},  the potential should be exponential
\begin{equation}\label{pot3}
V( T) \simeq e^{- T} .
\end{equation}
For definiteness  we will assume that the potential $V( T)$ in the Born-Infeld type action
(\ref{action4}) is a smooth  function interpolating between two
asymptotic expressions,   (\ref{pot1}) or  (\ref{pot2})    at maximum 
and (\ref{pot3}) at infinity.

The main  conclusions of our paper will be valid for the theories with the actions (\ref{action4}), (\ref{action1}),
(\ref{action5}), as well as with the alternative actions proposed in  \cite{alt1,alt2}.  
However, one should use all of these expressions with some care.  Rigorous derivation of the tachyon action have been performed in the approximation
truncated up to the second order derivatives of $\partial_{\mu} T$.
Since the tachyonic mass squared is $O(1)$ in units of $M_s^2$, all powers of $\partial_{\mu} T$ are important for its evaluation. Therefore the
truncated approximations in general do not
reproduce the correct values of the
tachyon masses $m^2=-M_s^2$ and $m^2=-M_s^2/2$ for
bosonic and supersymmetric cases \cite{Polchinski}.\footnote{We are grateful to
David Kutasov for clarification of this point.}

\section{Domains and Defects}\label{sec:defects}

It would be natural to start the investigation of the  tachyon condensation with 
tachyon at the top of its potential, i.e. at the moment before the condensation began.
Here we would like to make a simple comment concerning this possibility. 
Consider unstable non-BPS D3 brane with a  real tachyon field $T$.
Suppose for the moment that the spacetime geometry is flat.
If initially the tachyon field were absent, $T = 0$, then spontaneous 
symmetry breaking (due to  quantum fluctuations)
 and generation of the  tachyon condensate would divide  the 
universe into many small domains containing either positive or negative 
field $T$.
 In the superstring case the potential $V(T)$ is symmetric, see Eq. (\ref{pot2}), 
 and the universe always remains divided into domains of 
different types, separated by
domain walls.
 The resulting structure is incompatible with 
cosmological observations.

There are several ways to resolve this problem. First of all, one may 
try to find  models where the minimum of $V(T)$ at $T>0$ is deeper than the one 
at $T<0$. Then the domain walls will eventually disappear. It may also happen that initially in  our part of the universe the 
field $T$ was displaced from the top and 
 was, say,  positive. This would require the existence of a preceding  
stage of inflation preparing asymmetric but homogeneous initial 
conditions for the field $T$ in our patch of the universe. One could achieve these goals in  models 
based on  hybrid inflation, but this is a very non-trivial task, see in 
this respect \cite{Lazarides:2002zz} and references therein.

A simpler possibility would be to have inflation near $T = 0$, just like 
in the new inflation scenario. This is possible if $|m^2| \ll H^2$ near the top of the 
effective potential. Inflation in such models is eternal 
because of quantum fluctuations occasionally returning the field to the 
top of the effective potential \cite{Vilenkin:xq,Linde:fc}, and also 
because of the eternally inflating topological effects 
\cite{Linde:1994hy,Vilenkin:1994pv,Linde:1994wt}.

Inflation in the context of the theory of  tachyon condensation
proposed in \cite{Fairbairn:2002yp,Choudhury:2002xu},
is just a realization of this inflationary
scenario. Thus, to avoid the domain wall problem in  tachyon matter cosmology,
we need  to introduce inflation in the model.
We will examine this possibility in the next section.

Now consider hybrid-type inflation with merging $D$--$\bar D$ branes.
In this case tachyon field is complex, the potential has a phase symmetry
$V(T)=V(e^{i \alpha}T)$ \cite{sen02}. In this model
one would  expect formation of the cosmic strings  after annihilation of branes. 
However, the situation is more complicated if indeed the ground state of the tachyon
lies at $T \to \infty$. The string formation in this case is a long-lasting
process. It continues  as the field $T$ rolls
towards infinity. There will be one-dimensional regions of space (strings)
where $T=0$ and energy density is $\tau_3$. But even outside of these regions the energy density will be large due to gradients of the field $T$.
If $\tau_3$ is large, the distribution of energy density associated with strings of this type leads to very large inhomogeneities, which tends to make the models of such type 
 incompatible with cosmology; see also Section 5 for a discussion of other cosmological problems in the theories with $V(T)$ having a minimum at $|T| \to \infty$.

\section{Problems with  Tachyon Inflation}\label{parameters}

In this section we will study the possibility to achieve tachyonic inflation.
Assuming that inflation occurs near the top of the tachyon potential, we have
\begin{equation}\label{hubbleNEW}
H^2 = {\tau_3\over 3 M_p^2} = {g_s M_s^2 \sqrt 2 \over 3(2\pi)^3 v } \ .
\end{equation}
Inflation near the top of the tachyon potential is possible if $H^2 \gg 
|m^2|$.
This leads to the  condition 
\begin{equation}\label{inflcond}
 g_s \gg {3(2\pi)^3\over \sqrt 2} v \sim 260 v  \ .
\end{equation}
for the supersymmetric case, with $|m^2| = M_3^2/2$. 
For bosonic theory the {\it r.h.s.} value of this inequality is doubled.

If this condition is satisfied, inflation continues even when the field $T$ moves rather far from $T=0$, so that the exponential factor $e^{-T^2/4}$ becomes much smaller than 1. However, because of a very strong dependence of this factor on $T$, the end of inflation still occurs at $T = O(1)$. Keeping this in mind, let us estimate the value of $H$ at the end of inflation.

The equation for the time-dependent rolling tachyon in an expanding
universe follows from (\ref{action4}). It will be useful to recover the dimensions of the 
tachyon in units of $M_s$
\begin{equation}\label{time}
\frac{\ddot T }{1-\alpha' \dot T^2}+ 3H\, \dot T+ M_s^2\, \frac{V'}{V} = 0\ .
\end{equation}

One of the conditions for inflation is $|\dot H| \ll H^2$.
 The meaning of this condition is that the Hubble constant does
 not decrease much during a typical time $H^{-1}$. During inflation 
\begin{equation}\label{hubble1}
H^2 = {V(T)\over 3 M_p^2} = {g_s^2 V(T) \over 3v M_s^2} \ .
\end{equation}
Thus,
\begin{equation}\label{hubble2}
\dot H  =   {g_s^2 V'(T) \over  6 v M_s^2 H}\dot T \ .
\end{equation}
For definiteness, here and below we will discuss the
supersymmetric theory. The slow-roll equation for the field $T$ in the theory with the potential (\ref{pot2}) is
\begin{equation}\label{hubble3}
3H \dot T  =   {V'(T) M_s^2 \over 4 V \, \ln 2\  } \ .
\end{equation}
As a result, the condition $H^2\gg |\dot H|$ reads
\begin{equation}\label{hubble4}
H^2 \gg {(V')^2 M_s\over 24 V^2 \ln 2} \ .
\end{equation}
For $T = O(1)$ and $m^2 = -{M_s^2\over 8 \ln 2 }$ this condition takes a very simple form
\begin{equation}\label{hubble5}
H  \gg  |m| \ .
\end{equation}
Thus, at the end of inflation one has $H \sim |m|$. To derive this result we used a particular form of the potential assuming the Born-Infeld type action for the tachyon field. If we use a different action, some of the numerical factors in our equations will change. However, we expect that the general conclusion that $H \sim |m|$ at the end of inflation will remain valid. Indeed, this simple result  is valid in most of the versions of inflationary cosmology. 

The amplitude of gravitational waves produced during  inflation is
$h \sim {H\over  M_p}$.   Observational
 constraints on CMB anisotropy imply that at the end of inflation, when $H \sim |m|\sim M_s/\sqrt2$, one should have
\begin{equation}\label{hubble6}
{H\over M_p}  \leq  3.6\times 10^{-5} \ .
\end{equation}
This implies that
\begin{equation}\label{hubble7}
{M_s^2\over M_p^2} \sim {g_s^2\over v} \leq 2\times 10^{-9} \ .
\end{equation}
Thus, one should have
\begin{equation}\label{hubble8}
 g_s^2 \leq  2\times 10^{-9} v \ .
\end{equation}
One could find a similar result by investigation of density perturbations produced by the tachyon field, but
the investigation of gravitational waves is much simpler to perform.

This inequality could be compatible with the inflationary condition (\ref{inflcond}) only if 
\begin{equation}\label{hubble9}
v\ll 10^{-13} \ .
\end{equation}
 Meanwhile, as we explained, effective 4D theory is applicable only if $v\gg 1$. This means that even if inflation
 is possible in the models considered above, it cannot describe the formation of our part of the universe.
 To avoid huge anisotropy of the CMB, one should have a subsequent long stage of inflation produced by a different mechanism. 

This conclusion is not entirely unexpected because in the models we are considering now, unlike in the models  considered in \cite{renata} - \cite{Sarangi:2002yt}, we do not have small parameters that could be responsible for the extremely small anisotropy of the cosmic microwave background radiation. Understanding of this result requires some familiarity with inflationary cosmology and the theory of gravitational wave production. However, one can look at this situation from a different perspective. 

First of all, the inflationary condition $H \gg |m|\sim M_s$ implies that the wavelength of inflationary perturbations $\sim H^{-1}$ is of the same order  or less 
 than $l_s$. This is the first warning sign, suggesting that one cannot  study tachyonic inflation in the context of
an  effective 4D theory describing tachyon condensation.

Secondly, let us compare the brane tension $\tau_3$ with the Planck density $M_p^4$:
\begin{equation}\label{tension1}
{\tau_3\over  M_p^4} =  {g_s^3 \over {(2\pi)^3 v^2}} \ ,
\end{equation}
Comparing it with the inflationary condition $g_s \gg {3(2\pi)^3\over 2} v$, we find 
\begin{equation}\label{tension2}
{\tau_3\over  M_p^4} =  {27(2\pi)^6 \over {8}}v \ .
\end{equation}

Thus $\tau_3 \gg M_p$ if $v>1$, as we assumed. More explicitly, for superstring theory ($d=6$) one finds
\begin{equation}\label{tensionNEW}
{\tau_3\over  M_p^4} =  {27 \over {8 \pi}}\, \left({2\pi r \over l_s}\right)^6 \ .
\end{equation}
which means that $\tau_3 \gg M_p^4$ not only if one assumes that the radius of compactification is greater than the string length, $r\gg l_s$,  but even if one  makes a mild  assumption that $2\pi r \gg l_s$.

This suggests that 4D gravity is inapplicable for the description of inflation due to tachyon condensation near the top of the effective potential $V(T)$. Meanwhile, Eq. (\ref{hubble9}) shows that even if we forget about it and continue calculations, we will find that the tachyonic inflation would produce  metric perturbations of acceptable magnitude only if $r \ll l_s$.

However, let us try to take a positive attitude to the situation. The constraints related to gravitational wave production
derived   in 4D theory are very difficult to overcome. 
But this just means that  tachyonic inflation can not be responsible for the last 
60 e-folds of inflation. This last stage of inflation determined the large scale structure of the observable part of the universe.
  It might be possible, however, that  tachyonic inflation is responsible for an earlier stage of inflation. The existence of this stage may be important for the resolution of major cosmological problems such as the homogeneity, flatness and isotropy problems. This could be a `bad' inflation, a very short one, producing large density perturbations. But one can always produce density perturbation at a later stage of inflation driven by some other mechanism. Meanwhile, the best way to resolve the problems mentioned above is to have inflation that could begin directly at the Planck density \cite{Chaot,book}. 
In this sense, the Planck density tachyonic inflation could be an interesting possibility.
Also, the constraint that follows from investigation of the gravitational wave production may be very different in higher dimensional theories with brane inflation,
in particular because the effective 4D Planck  mass may  depend on the warp geometry.
Investigation of this possibility requires a very different approach.

In this respect, depending on one's attitude, one may consider tachyonic inflation either marginally impossible, or marginally possible. Indeed, if we do not need to achieve 60 e-folds of inflation, we do not need to have $H \gg m$.
 Similarly, our conclusion that inflation requires super-Planckian density, ${\tau_3\gg  M_p^4}$ should be examined more carefully.
 String theorists use a `small' Planck mass $M_p \sim 2.4 \times 10^{18}$ GeV, whereas quantum gravity fluctuations become large at $V(T) > \bar M_p^4$ where $\bar M_p = \sqrt {8\pi} M_p$. As a result, the `gravitational' Planck density is $(8\pi)^2 \sim 600$ times larger than the `stringy' Planck density considered in our investigation. 

Therefore one should not discard the possibility that whereas tachyon condensation probably is not responsible for the last 60 e-folds of inflation,  it could be responsible for the first stage of inflation, solving all major cosmological problems. Eq. (\ref{tensionNEW}) does not offer much support to this possibility, but one cannot completely rule it out either.

\section{Problems with Early Matter  Domination}\label{sec:radiation}

We have seen above that it is very unlikely that the last 60 e-folds of inflation appear due to a
non-BPS brane tachyon rolling from the top
of its potential. In this section we give another,
qualitatively different argument, why
  models with the potentials (\ref{pot1})-(\ref{pot3})
cannot describe the last  60 e-folds of inflation responsible for the large-scale structure formation and creation of matter in the universe.
This argument is also valid for another situation, when
brane hybrid inflation is supported by inter-brane separation
(for a $D$--$\bar D$ system)
and the tachyon  
rolls to its ground state at $T \to \infty$
immediately after the brane-antibrane annihilation.

According to the inflationary  scenario discussed in the previous section,
 the energy density of the tachyon 
field dominated the total energy density of the universe during 
inflation. After the end of inflation 
the tachyon field with the potential $e^{-T}$ at large $T$
almost immediately acquires  a matter equation of state $\varepsilon_T \sim a^{-3}$.
 This means 
that the energy density of all other kinds of matter, including 
radiation $\varepsilon_r \sim a^{-4}$,
 were falling down much faster than the energy density of the 
tachyon field.

Moreover, after inflation there were no other types of matter. All 
matter that appears after inflation should be created in the process of 
(p)reheating. In most versions of the theory of reheating, production of 
particles occurs only when the inflaton field oscillates near the 
minimum of its effective potential.
 However, the effective potential of 
the rolling tachyon in theories (\ref{action1}), (\ref{action5})
does not have any minimum at finite $T$, so this
mechanism does not work. 
There exists a mechanism of reheating due 
to  gravitational particle production. 
 However, this mechanism is extremely 
inefficient at $H \ll M_p$.
Even if a small fraction of tachyon energy  $\varepsilon_T$ is released into
radiation, it will be rapidly redshifted away.
Thus, in this scenario the energy density of the tachyon field {\it
always} dominates the universe.
This is incompatible with the well 
known fact that at some early stage (in particular, during 
nucleosynthesis) our universe was radiation dominated. The only way to resolve this problem is to assume that there was a separate stage of inflation at the time when the tachyon field energy was exponentially small. But this would require fine-tuning \cite{Shiu:2002qe} and imply that the tachyon played no role in the  early universe until the very late stages of its evolution.

\section{Summary}\label{sec:summ}

We conclude that
 string tachyon condensation might be responsible for a short stage of inflation at a nearly Planckian density, thus solving the major cosmological problems. However, this possibility is 
 at the verge of applicability of 
the effective 4D theory used to describe  tachyon condensation. The main reason is that the inflationary condition $H\gg |m|$ is satisfied only for  super-Planckian energy densities (D3-brane tension), under the standard condition that the radius of compactification is greater than the string length. Moreover, it is very unlikely that the   tachyon condensation studied in \cite{sennonbps} can be responsible for the last 60 e-folds of inflation, because in order to produce perturbations of the metric smaller than $10^{-4}$ one needs to have an energy density at least 9 orders of magnitude smaller than the Planck density.  
These conclusions do not depend on a detailed choice between various formulations of  string tachyon theory (whether one studies only the theory as it is formulated, say, in \cite{shatash,Kutasov1,Kutasov2} or  the Born-Infeld-type form   
\cite{Sen1} or  in alternative forms \cite{alt1,alt2}).

One may also consider  inflationary models of the hybrid inflation type, where inflation is driven by some other mechanism, but the end of inflation occurs due to tachyon condensation.  We have found that if the tachyon potential
in these theories has a minimum at $T\to \infty$, as in \cite{sennonbps}, then the universe in such models never becomes radiation dominated, in contradiction with the theory of nucleosynthesis. One could resolve this problem if there was a  separate stage of inflation at the time when the tachyon  energy was fine-tuned to be exponentially small. But this would  imply that the tachyon played no role in the post-inflationary  universe until the very late stages of its evolution. Thus it seems that the tachyon condensate can be a  dark  matter candidate only in the context of rather complicated and fine-tuned cosmological models.

 Alternatively, one may consider models of hybrid inflation 
\cite{Hybrid} where brane cosmology leads to the formation of a tachyon 
condensate of a different type \cite{stringyhybrid} - 
\cite{Sarangi:2002yt}. This condensate can be represented by a complex 
field $\phi$. In some of these models the shape of the tachyonic effective potential 
$V(\phi)$ has been explicitly calculated, see e.g.  \cite{renata,renata2}.  
 To achieve  a successful cosmological scenario, this potential, unlike the potentials 
for the tachyon condensate considered in \cite{sennonbps,shatash,Kutasov1,Kutasov2,Choudhury:2002xu,Sen1},
 should have a minimum not at $\phi \to \infty$ but  at $|\phi| \ll M_p$  \cite{renata} - \cite{Sarangi:2002yt}.

We are grateful to G. Dvali, R. Kallosh, D. Kutasov and A. Sen for useful comments,
and G. Felder for help.
 The work by L.K. was supported by  NSERC and  CIAR. The work by A.L. was supported
by NSF grant PHY-9870115, and by the Templeton Foundation grant
No. 938-COS273.  L.K. and A.L. were also supported
  by NATO Linkage Grant 97538.

\end{document}